\documentclass[journal]{IEEEtran}
\usepackage[cmex10]{amsmath}
\usepackage{graphicx}
\usepackage{cite}
\usepackage{epstopdf}
\usepackage{algorithm}
\usepackage{algorithmic}
\usepackage{amsfonts}

\begin{document}
\title{Energy-efficient resource allocation for hybrid bursty services in multi-relay OFDM networks}
\author{Yuhao~Zhang, Qimei~Cui,~\IEEEmembership{Senior~Member,~IEEE,} Ning~Wang, Yanzhao~Hou, and Weiliang~Xie
\IEEEcompsocitemizethanks{
\IEEEcompsocthanksitem The work was supported by National Nature Science Foundation of China Project (Grant No. 61471058), Hong Kong, Macao and Taiwan Science and Technology Cooperation Projects (No. 2014DFT10320, No. 2016YFE0122900), Beijing Nova Program (No. xx2012037), Shenzhen Science and Technology Project (No. 20150082), and Beijing Training Project for The Leading Talents in S\&T (No. Z141101001514026).
\IEEEcompsocthanksitem Y. Zhang, Q. Cui, N. Wang, and Y. Hou are with the School of Information and Communication Engineering, Beijing University of Posts and Telecommunications, Beijing 100876, China (e-mail: cuiqimei@bupt.edu.cn).
\IEEEcompsocthanksitem W. Xie is with the China Telecom Corporation Limited, Beijing 100140, China (e-mail: xiewl.bri@chinatelecom.cn).}}

\maketitle
\begin{abstract}
In this paper, we propose Mapped Two-way Water Filling~(MTWF) scheme to maximize Energy Efficiency~(EE) for hybrid bursty services with Quality of Services~(QoS) requirements in Two-Way Multi-Relay~(TWMR) OFDM networks. The bursty traffic is first analyzed by strictly proved equivalent homogeneous Poisson process, based on which the QoS requirements are converted into sum-rate constraints. The formulated non-convex EE maximization problem, including subcarrier assignment, Relay Selection~(RS) and rate allocation, is NP-hard involving combinatorial optimization. To conduct optimal RS on each subcarrier without priori bursty traffic knowledge, we utilize some approximate relationships under high data rate demands to remove its dependence on two-way data rates, and simplify the whole optimization problem as well. After the optimal channel configuration is obtained, which only depends on channel conditions, subcarrier assignment is attained through Elitist Selection Genetic Algorithm~(ESGA), and rate allocation of each service is fulfilled by deducing two-way water filling principle. A new equivalent optimization objective function is proposed next as the simple evaluating index in ESGA to reduce complexity. Finally, simulations are carried out to verify the superiority and convergence of our scheme, as well as the applicability for different scenarios.
\end{abstract}


\section{Introduction}\label{sec1}
Recently, it is widely acknowledged that the evolution of telecommunication industry leads to the rapid and intolerable increase of energy consumption all over the world~\cite{Ref1,Ref46}. Moreover, there is an exponentially growing gap between the energy demand of mobile equipments and the battery capacity due to the mismatch of larger device energy consumption and slower battery development~\cite{Ref2}. As a result, Energy Efficiency~(EE) in wireless networks has become more important, and obtained much more research efforts. Although many researches have been carried out towards some certain system indexes, e.g., minimal Bit-Error-Rate~(BER) and maximal Spectral Efficiency~(SE)~\cite{Ref45}, it is worth noting that the implement of these indexes cannot achieve the optimal EE at the same time. For example, it is proved that single Relay Selection~(RS) is the optimal scheme for energy efficiency under two-way relay channels with Decode-and-Forward~(DF) protocol~\cite{Ref4}, while the dual RS can obtain the optimal BER performance compared with other considered RS schemes~\cite{Ref3}.

The wireless relay node, deployed in the middle of two source nodes to assist the information transmission, was introduced in 3GPP-LTE-Advanced~\cite{Ref5} for throughput improvement and coverage extension without transmission power increase, particularly in cellular edge and signal blind area. Therefore, it can reduce the overall energy consumption of wireless networks significantly, but SE will be affected since more time slots are wasted with typical Half-Duplex~(HD) transmission~\cite{Ref6}. In order to overcome this deficiency, two-way relay with Physical-layer Network Coding~(PNC), which takes advantage of the additive nature of electromagnetic wave for equivalent network coding operation at physical layer, is proposed to compensate the spectral loss and halve the time slots~\cite{Ref6,Ref7}. At present, there are so many researches about two-way relay networks, e.g., transmission protocols~\cite{Ref8}, achievable rate regions~\cite{Ref9}, time assignment~\cite{Ref10} and Power Allocation~(PA)~\cite{Ref11,Ref44}. Applying two-way relay into OFDM networks with PNC to obtain optimal EE is much more interesting, which has attracted considerable research attentions so far. In~\cite{Ref14}, considering fixed and dynamic circuit power, the strict quasi-concave energy-efficient power allocation problem is resolved by Dinkelbach's method for fractional programming under two-way Amplify-and-Forward~(AF) relay networks. In~\cite{Ref15}, the active subcarriers and its allocated data bits are jointly optimized for transmission power minimization, where the bidirectional water filling is proposed with the balance of transmission and circuit power. In~\cite{Ref16}, to prolong the lifetime of battery-operated AF relay, the mixed-integer nonlinear EE programming problem, w.r.t relay selection and power loading, is resolved by exploiting energy pricing concept and max-min energy efficiency criterion. In addition, the transmission duration optimization for EE is considered under MIMO OFDM networks~\cite{Ref17}.

Under multi-relay networks, it is known that the joint RS and PA scheme is an efficient method to promote system performance~\cite{Ref16,Ref17,Ref42,Ref12,Ref13}, which has been mainly studied in this paper. In~\cite{Ref4}, the joint RS and PA scheme is proposed to minimize the energy consumption of Analog Network Coding (ANC) under given end-to-end data rate requirement. It is proved that single RS is most energy-efficient, based on which the closed-form expression of the optimal transmission power was derived. Moreover, the joint RS and PA scheme is adopted to improve the Symbol Error Probability~(SEP) performance~\cite{Ref12}, the received Signal-to-Noise-Ratios~(SNRs)~\cite{Ref13} and the system EE~\cite{Ref42,Ref43}. Unfortunately, compared with single relay networks, there are less relevant works on EE optimization in Two-Way Multi-Relay~(TWMR) OFDM networks. In~\cite{Ref18}, the joint power and subcarrier allocation, proved to be a quasi-concave fractional programming problem, is resolved by Dinkelbach's method. In~\cite{Ref19}, considering linearly rate-dependent circuit power, the EE optimization problem is resolved approximately under proportional rate constraints in DF relay beamforming networks by jointly optimizing relay selection, subcarrier assignment and power allocation. Furthermore, as complex MIMO channel is introduced in~\cite{Ref20}, the EE optimization problem is proved to be concave, which is resolved by employing dual decomposition approach. In~\cite{Ref21}, the optimal EE resource allocation is proposed under the constraints of required data rates, where assigning each subcarrier to a unique relay is proved to be the best energy-efficient scheme. However, equal rate allocation is adopted in this paper to reduce complexity, may causing sub-optimal results in practice.

These works above do not consider common and universal hybrid services and its corresponding cooperation. Unlike single service, hybrid services have to share all the limited transmission resources in networks. Moreover, data packets in practical system arrival exhibits statistic bursty characteristic and correlated feature. In terms of bursty traffic, there are many researches on internet and computer science~\cite{Ref22,Ref23,Ref24}, but much less works are found in wireless networks. In~\cite{Ref25}, scheduling policies that are robust to bursty traffic are designed by constituting a queue system with two conflicting links, where the first link is modeled by heavy-tailed arrival process and the second link is modeled by light-tailed one. In~\cite{Ref26}, considering the bursty traffic impact on Base Station~(BS) sleeping, the total power consumption and average delay for bursty traffic, which follows the Interrupted Poisson Process~(IPP), are analyzed. In~\cite{Ref27}, closer to the real hybrid networks, a new equivalent analytical model is developed to investigate the performance indexes (including end-to-end delay, loss probability and throughput), which are able to capture the bursty and correlated features. To the best of our knowledge, energy-efficient resource allocation for hybrid bursty services has been never studied in TWMR OFDM networks.

In this paper, we proposed Mapped Two-way Water Filling~(MTWF) scheme to maximize EE for hybrid bursty services with Quality of Services~(QoS) requirements in TWMR OFDM networks. The bursty traffic is first analyzed by queue theory, where an strictly proved equivalent homogeneous Poisson process (Poisson flow) is established, based on which the QoS requirements can be converted into sum-rate constraints through the superposition of independent Poisson flows. For each subcarrier, optimal single Relay Selection~(RS) is obtained, but its dependence on two-way rate demands will make the whole problem very complicated to resolve. Therefore, approximate relationships under high data rate demands are introduced to simplify RS, based on which optimal channel configuration can be attained only using channel information. After optimal RS, the original non-convex optimization problem is transformed into equivalent NP-hard combinatorial problem without closed-form analytic solutions~\cite{Ref15}. Then, the optimal subcarrier assignment is resolved by Elitist Selection Genetic Algorithm~(ESGA), where an equivalent objective function is further proposed to simplify the evaluating index, followed by the optimal rate allocation, which is obtained by deducing two-way water filling principle. Finally, simulations are carried out to demonstrate superiority and efficiency of our scheme for both symmetric and asymmetric traffic.

The contributions of this paper are summarized as follows.
\begin{itemize}
  \item analyse hybrid bursty services precisely, and create an strictly proved equivalent queue system to convert QoS requirements into sum-rate constraints, which is key to formulating EE optimization problem.
  \item decompose the original non-convex NP-hard optimization problem into three subproblems. Using some reasonable approximations, the global optimal closed-form solutions are derived for the relay selection and rate allocation problem, and the low-complexity heuristic ESGA, keeping the best solution after roulette selection, is designed to solve the subcarrier assignment problem.
  \item reveal that our scheme, having a proposed equivalent objective function to simplify the evaluating index in ESGA, can get the optimality and the convergence with polynomial computational complexity by theoretical analysis and Monte Carlo simulations.
\end{itemize}

The rest of the paper is organized as follows. System model is described in Section~\ref{sec2}. EE maximization problem is formulated in Section~\ref{sec3} and then resolved in Section~\ref{sec4}. In Section~\ref{sec5}, we present the equivalent optimization objective function, followed by algorithm implementation and analysis in Section~\ref{sec6}. Simulation results are provided in Section~\ref{sec7}, followed by the conclusion in Section~\ref{sec8}.

\section{System model}\label{sec2}
\subsection{Scenario Description}\label{subsec2:1}
\begin{figure}[!t]
\centering
\includegraphics[scale=0.3]{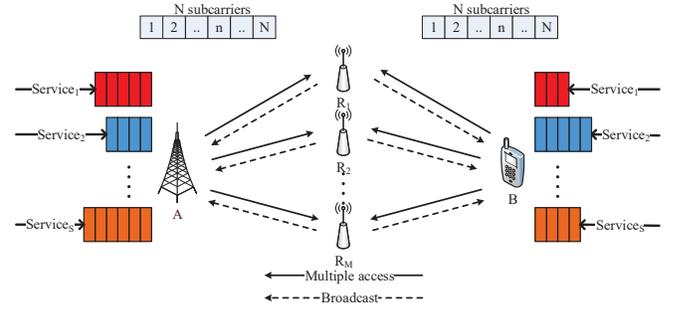}
\caption{Hybrid bursty services transmission in TWMR OFDM networks}
\label{pic1}
\end{figure}

The system we considered contains a base station~(A), a mobile user~(B) and $M$ relay nodes, as shown in Fig.~\ref{pic1}. The channel condition between A and B is too weak to communicate directly due to some limitations, e.g., far distance or obstacles. Therefore, the information exchange can only be fulfilled with the assistance of $M$ relays, which are supposed to adopt PNC with AF protocol. It is assumed that all nodes are equipped with single antenna and operate in HD mode. Perfect channel modulation and coding schemes are assumed to get the Shannon capacity, which make the power consumption minimum.

There are $N$ subcarriers in the system, where the whole frequency band is divided into mutually disjoint narrow-band subcarriers with the same fixed bandwidth by OFDM. It is assumed that the channel coherence bandwidth is larger than the bandwidth of individual subcarrier, which experiences different frequency-flat fading due to frequency selectivity. According to channel reciprocity of Time Division Duplex~(TDD), the channel coefficient from A to B is the same as that from B to A. Furthermore, it is also assumed that all nodes can perfectly detect and predict the channel conditions during the time frame. Considering large scale path loss and Rayleigh fading effect, the channel coefficient from A and B to relay $m$ on subcarrier $n$ are denoted by $h_m^n$ and $g_m^n$, respectively. And $n_n$ is the additive white Gaussian noise at all nodes on subcarrier $n$ with power $\sigma_n^2$.

\begin{figure}[!t]
\centering
\includegraphics[scale=0.4]{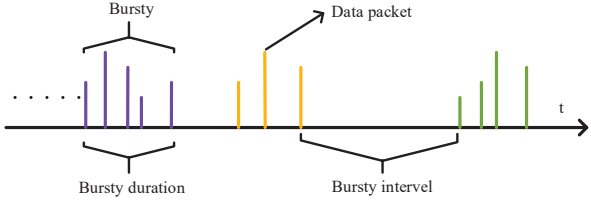}
\caption{Bursty traffic of single service}
\label{pic_Bursty}
\end{figure}

There are $S$ services that have to be transmitted between A and B with different characteristics and QoS requirements. It is assumed that each service has the same max-delay but different arrival rate and average packet length in two directions. The max-delay of these $S$ services are denoted by $D_1$, $D_2$, $D_3$, $\cdots$, $D_S$. As shown in Fig.~\ref{pic_Bursty}, it is assumed that the bursty traffic arrives in the form of Poisson process and its duration time satisfies negative exponential distribution. And during its duration, the data packet arrival is regarded as the Poisson process and the length of the packet also obey Poisson distribution. We denote the arrival rate of bursty traffic of these $S$ services by $\Lambda_1^1$, $\Lambda_1^2$, $\Lambda_1^3$, $\cdots$, $\Lambda_1^S$ and $\Lambda_2^1$, $\Lambda_2^2$, $\Lambda_2^3$, $\cdots$, $\Lambda_2^S$ in two directions, respectively. The parameters of the negative exponential distribution of bursty duration are given by $\frac{1}{T_1^1}$, $\frac{1}{T_1^2}$, $\frac{1}{T_1^3}$, $\cdots$, $\frac{1}{T_1^S}$ and $\frac{1}{T_2^1}$, $\frac{1}{T_2^2}$, $\frac{1}{T_2^3}$, $\cdots$, $\frac{1}{T_2^S}$. The arrival rate and the average length of data packet during the bursty duration for these $S$ services are donated by $\lambda_1^1$, $\lambda_1^2$, $\lambda_1^3$, $\cdots$, $\lambda_1^S$ and $L_1^1$, $L_1^2$, $L_1^3$, $\cdots$, $L_1^S$ from A to B, while $\lambda_2^1$, $\lambda_2^2$, $\lambda_2^3$, $\cdots$, $\lambda_2^S$ and $L_2^1$, $L_2^2$, $L_2^3$, $\cdots$, $L_2^S$ from B to A, separately. The infinite data buffer is assumed to support all the packets without any loss.

It is noted that only one mobile user case is studied in this paper. However it can represent the advanced multiple users case by gathering all the services together to form a virtual user on some situations. Therefore, our scheme proposed in this paper can be used in multi-user case if the following conditions are satisfied: 1) the base station can coordinately manage and allocate resource for all the users; 2) any subcarrier cannot be reused or shared, i.e., one subcarrier can only be allocated to one user.
\subsection{System Capacity}\label{subsec2:2}
It is obvious that PNC requires two time slots to complete information exchange, i.e., Multiple Access Phase~(MAC) and Broadcast Phase~(BC). In the first phase, node A and B transmit information $x_{1,s}^n$ and $x_{2,s}^n$ with transmission power $P_{A,s}^n$ and $P_{B,s}^n$ to the $M$ relays on subcarrier $n$ for service $s$, respectively. With perfect synchronization, the received information at relay $m$ can be obtained, as given by
\begin{equation}\label{Equ1}
    y_{s,m}^n = h_m^n\sqrt{P_{A,s}^n}x_{1,s}^n + g_m^n\sqrt{P_{B,s}^n}x_{2,s}^n + n_n.
\end{equation}

If relay $m$ is selected to take part in the transmission on subcarrier $n$ for service $s$, indicated by $\rho_{m,s}^n$, it will forward the received information according to AF protocol and broadcast with transmission power $P_{R,m,s}^n$. It is noted that $\rho_{m,s}^n = 1$ if relay $m$ is selected to take part in the transmission on subcarrier $n$ for service $s$, otherwise $\rho_{m,s}^n = 0$. The amplification factor at relay $m$ can be written as
\begin{equation}\label{Equ2}
    \sqrt{\alpha_{R,m,s}^n} = \sqrt{\frac{P_{R,m,s}^n}{|h_m^n|^2P_{A,s}^n + |g_m^n|^2P_{B,s}^n + \sigma_n^2}}.
\end{equation}
Therefore, the received information at A and B from all selected relays can be obtained, as given by
\begin{equation}\label{Equ3}
    Z_{A,s}^{n} = \sum_{m=1}^{M}\rho_{m,s}^nh_m^n\sqrt{\alpha_{R,m,s}^n}y_{s,m}^n + n_n,
\end{equation}
\begin{equation}\label{Equ4}
    Z_{B,s}^{n} = \sum_{m=1}^{M}\rho_{m,s}^ng_m^n\sqrt{\alpha_{R,m,s}^n}y_{s,m}^n + n_n.
\end{equation}
Since node A and B know its own transmitted information, by perfect self-interfere cancellation in the received signal, the desired information can be obtained at A and B, as given by
\begin{equation}\label{Equ5}
    Z_{A,s}^{n^*} = \sum_{m=1}^{M}\rho_{m,s}^nh_m^n\sqrt{\alpha_{R,m,s}^n}\left(g_m^n\sqrt{P_{B,s}^n}x_{2,s}^n + n_n\right) + n_n,
\end{equation}
\begin{equation}\label{Equ6}
    Z_{B,s}^{n^*} = \sum_{m=1}^{M}\rho_{m,s}^ng_m^n\sqrt{\alpha_{R,m,s}^n}\left(h_m^n\sqrt{P_{A,s}^n}x_{1,s}^n + n_n\right) + n_n.
\end{equation}
According to the SNRs and Shannon capacity formula, the downlink (from A to B) and uplink (from B to A) achievable end-to-end data rates $r_{1,s}^n$ and $r_{2,s}^n$ can be formulated, as given by
\begin{equation}\label{Equ9}
    r_{1,s}^n = \frac{1}{2} \cdot \frac{W}{N} \cdot \log\left(1 + \frac{\sum_{m=1}^{M}\rho_{m,s}^n|h_m^n|^2|g_m^n|^2\alpha_{R,m,s}^nP_{A,s}^n}{\sigma_n^2 + \sum_{m=1}^{M}\rho_{m,s}^n|g_m^n|^2\alpha_{R,m,s}^n\sigma_n^2}\right),
\end{equation}
\begin{equation}\label{Equ10}
    r_{2,s}^n = \frac{1}{2} \cdot \frac{W}{N} \cdot \log\left(1 + \frac{\sum_{m=1}^{M}\rho_{m,s}^n|h_m^n|^2|g_m^n|^2\alpha_{R,m,s}^nP_{B,s}^n}{\sigma_n^2 + \sum_{m=1}^{M}\rho_{m,s}^n|h_m^n|^2\alpha_{R,m,s}^n\sigma_n^2}\right),
\end{equation}
where $W$ is the system bandwidth. Note that the pre-log factor $1/2$ comes from the two time slots required to information exchange.
\section{Problem Formulation}\label{sec3}
In this section, we will first discuss the QoS requirements of hybrid bursty services. Without loss of generality, only downlink is analyzed here, and the uplink can be obtained through the same way. Form \eqref{Equ9}, the downlink transmission rate of service $s$ on subcarrier $n$ is $r_{1,s}^n$~($n$=1,2,3,$\cdots$,$N$), where $r_{1,s}^n=0$ if subcarrier $n$ is not assigned to service $s$. Therefore, for service $s$, the data packet departure on subcarrier $n$ can be regarded as a homogeneous Poisson process with the parameter of $\mu_{1,s}^n=\frac{r_{1,s}^n}{L_1^s}$. The bursty arrival is the Poisson process with the parameter of $\Lambda_1^s$, thus the interval duration meets the negative exponential distribution, i.e., $f(t)=\Lambda_1^s \cdot e^{-\Lambda_1^s \cdot t},(t\geq0)$, where the average interval duration is $\frac{1}{\Lambda_1^s}$. And the mean value of the bursty duration for service $s$ is $T_1^s$. Therefore, the condition probability density function of $k$ arrived packets during bursty duration $t_B$ can be formulated, as given by
\begin{equation}\label{Equ_con_pro}
    P(k \mid t_B)=\frac{(\lambda_1^s \cdot t_B)^k}{k!}e^{-\lambda_1^s \cdot t_B},\ (k=0,1,2,\cdots).
\end{equation}

It can be obtained that the random variable $E(k \mid t_B)$ is equal to $\lambda_1^s \cdot t_B$, where $t_B$ is a random variable and $E(\cdot)$ represents the mean value of the corresponding random variable. Therefore, it is known that the average arrival data packets $E(k)$ is equal to $\lambda_1^s \cdot T_1^s$ since $E(k)=E[E(k \mid t_B)]$. Considering the fact that there is no data packet arriving during the bursty interval time, the average data packets arrived per one unit time can be formulated, as given by
\begin{equation}\label{Equ_Eq_Arrival_Rate}
   {\lambda_1^s}^*=\frac{\lambda_1^s \cdot T_1^s}{T_1^s + \frac{1}{\Lambda_1^s}}.
\end{equation}

It is clearly seen that the packet arrival during the whole time is a heterogeneous Poisson process, but a new simple homogeneous Poisson arrival with the parameter of ${\lambda_1^s}^*$ can be introduced to replace the original heterogeneous one to guarantee the equivalent average performance~(delay characteristic). The rigid proof is presented in~{Appendix A}.

Since the transmission on subcarrier $n$ is independent with each other, according to the superposition property of independent Poisson flows, the departure process of service $s$ can be interpreted as a Poisson flow with the following leaving rate, as given by
\begin{equation}\label{Poisson_superposition}
   \mu_1^s=\sum_{n=1}^{N}\mu_{1,s}^n=\frac{\sum_{n=1}^{N}r_{1,s}^n}{L_1^s},\ \forall s.
\end{equation}

By these equivalences, the data packets of service $s$ will arrive at the node according to homogeneous Poisson process and will be transmitted by one processing unit with negative exponential distributed processing time. A newly coming packet needs to wait and be stored in the infinite data buffer until its former packets are all leaving. Therefore, the queue system now can be modeled as a standard $M/M/1$ queue (birth and death process), which has been widely researched and used in communication networks~\cite{Ref36,Ref38,Ref39,Ref40}. Considering the statistical delay characteristic, the QoS requirements can be expressed through the average delay formula of $M/M/1$ model~\cite{Ref37}, as given by
\begin{equation}\label{Equ11}
    Delay_1^s = \frac{1}{\frac{\sum_{n=1}^{N}r_{1,s}^n}{L_1^s} - {\lambda_1^s}^*} \leq D_s,\ \forall s.
\end{equation}

In order to formulate objective function and constraints of the optimization problem, the subcarrier assignment index $C_s^n$ is introduced. Let $C_s^n = 1$ if subcarrier $n$ is assigned to service $s$, otherwise $C_s^n = 0$. Considering the practical significance, an implicit constraint is obtained, i.e., if $C_s^n = 0$, $\rho_{m,s}^n = 0$ must be established. Note that each subcarrier can only be used by one service, i.e., $\sum_{s=1}^S C_s^n = 1$. By substituting $C_s^n$ into \eqref{Equ11}, the sum-rate constraints in two directions can be obtained, as given by
\begin{equation}\label{Equ15}
    \sum_{n=1}^N C_s^n \cdot r_{1,s}^n \geq R_s^1,\ \sum_{n=1}^N C_s^n \cdot r_{2,s}^n \geq R_s^2,\ \forall  s.
\end{equation}
where
\begin{equation}\label{sum_rate_con}
    R_s^1=\frac{L_1^s}{D_s} + {\lambda_1^s}^* \cdot L_1^s,\ R_s^2=\frac{L_2^s}{D_s} + {\lambda_2^s}^* \cdot L_2^s.
\end{equation}

Next, we will maximize the EE of the hybrid bursty services in TWMR OFDM networks with the QoS requirements, i.e., the sum-rate constraints $\{R_s^1\}$ and $\{R_s^2\}$. For service $s$, $R_s^1$ and $R_s^2$ are the downlink and uplink sum-rate constraints to satisfy the QoS requirements precisely.

The EE is defined as the number of overall data bits transmitted in both link directions over the energy consumed by all nodes, donated by $E_{total}$, within the transmission duration $T$~\cite{Ref15,Ref19}, as given by
\begin{equation}\label{EE_indicator}
\begin{aligned}
{\eta _E} &= \frac{\sum_{s=1}^S (R_s^1+R_s^2) \cdot T}{E_{total}} = \frac{\sum_{s=1}^S (R_s^1+R_s^2) \cdot T}{P_{total} \cdot T} \\
&= \frac{\sum_{s=1}^S (R_s^1+R_s^2)}{P_{total}},
\end{aligned}
\end{equation}
where the third equality indicates that the EE is equivalent to the ratio between the average total data rate for all services (in both directions) and the average total transmission power consumption at all the nodes, denoted by $P_{total}$~\cite{RefEEmetric}. It is noted that with high data rate demands the fixed circuit power consumption is small enough to be ignored compared with $P_{total}$, which needs to increase exponentially to support higher data rate according to Shannon capacity formula. To this end, given the QoS requirements for all services, i.e., $\sum_{s=1}^S (R_s^1+R_s^2)$, maximizing $\eta_E$ is equivalent to minimizing $P_{total}$.

Minimizing $P_{total}$ also facilitates maximizing the EE of hybrid bursty services in TWMR OFDM networks we considered. The reason is that maximizing the EE directly needs to solve the complex fractional programming optimization problem~\cite{Ref19}, which is indeed impossible to deal with efficiently due to the very complicated structure of $P_{total}$, where the transmission powers of all nodes couple jointly and cannot be expressed independently, as shown in \eqref{Equ9} and \eqref{Equ10}.

Next, we will investigate the total power consumption minimization problem for hybrid bursty services to achieve optimal EE~\cite{Ref4,Ref15,Ref21}. To realize the goal of minimizing $P_{total}$ for hybrid bursty services with QoS requirements, the general form of the optimization problem can be formulated, as given by
\begin{equation}\label{Equ18}
    {\rm Min}\ \sum_{s=1}^S \sum_{n=1}^N C_s^n \left(P_{A,s}^n + P_{B,s}^n + \sum_{m=1}^M \rho_{m,s}^n P_{R,m,s}^n\right), \tag{\textbf{P1}}
\end{equation}
\begin{equation}\nonumber
    {\rm S.t.}\ \sum_{n=1}^N C_s^n \cdot r_{1,s}^n \geq R_s^1,\ \sum_{n=1}^N C_s^n \cdot r_{2,s}^n \geq R_s^2,\ \forall  s,
\end{equation}
\begin{equation}\nonumber
    \sum_{s=1}^S C_s^n = 1,\ C_s^n \in \{ 0,1 \},\ \rho_{m,s}^n \in \{ 0,1 \},\ \forall n,m,s,
\end{equation}
\begin{equation}\nonumber
    0 \leq P_{A,s}^n,\ 0 \leq P_{B,s}^n,\ 0 \leq P_{R,m,s}^n,\ \forall n,m,s.
\end{equation}
It can be proved that the optimal solution must reach the equality of the sum-rate constraints just by reducing $\{r_{1,s}^n\}$ and $\{r_{2,s}^n\}$ to make it hold for energy saving~\cite{Ref10}.

It is obvious that the non-convex mixed discrete and continuous optimization problem~\eqref{Equ18}, with enormous binary integer variables $\{C_s^n\}$ and $\{\rho_{m,s}^n\}$, is NP-hard without analytical solution~\cite{Ref15,Ref17,Ref19}. To achieve the global optimum of~\eqref{Equ18}, all possible solutions $\{\{C_s^n\},\{\rho_{m,s}^n\},\{r_{1,s}^n\},\{r_{2,s}^n\}\}$ have to be checked, i.e., the computer exhaustive search is needed, which may result in very high computational complexity. Therefore, instead of searching the solution exhaustively, we resolve this complex problem in three steps, corresponding to three subproblems which will be discussed in detail.
\section{Mapped Two-way Water Filling Scheme}\label{sec4}
The key thought of our scheme is first obtaining the optimal channel configuration, based on which we employ ESGA to iterate subcarrier assignment and propose two-way water filling algorithm to conduct rate allocation.
\subsection{Optimal Relay Selection}\label{subsec4:1}
Here, we will first explore the optimal relay selection on subcarrier $n$ for a given result of subcarrier and rate allocation, donated by $\{\{C_s^n\},\{R_{1,s}^n\},\{R_{2,s}^n\}\}$. Without loss of generality, it is assumed that subcarrier $n$ has been assigned to service $s$, i.e., $C_s^n=1$. For service $s$, $R_{1,s}^n$ and $R_{2,s}^n$ are the allocated data rate on subcarrier $n$ in the downlink and uplink, respectively; in other words, subcarrier $n$ needs to achieve data rates $R_{1,s}^n$ and $R_{2,s}^n$ in the two directions with the help of optimal relays. Therefore, the optimal relay selection problem on subcarrier $n$ can be formulated, as given by
\begin{equation}\label{Equ24}
    {\rm Min}\ P_{A,s}^n + P_{B,s}^n + \sum_{m=1}^M \rho_{m,s}^n P_{R,m,s}^n, \tag{\textbf{P2}}
\end{equation}
\begin{equation}\nonumber
    {\rm S.t.}\ r_{1,s}^n = R_{s,n}^1,\ r_{2,s}^n = R_{s,n}^2,
\end{equation}
\begin{equation}\nonumber
    \rho_{m,s}^n \in \{ 0,1 \},\ \forall m,
\end{equation}
\begin{equation}\nonumber
    0 \leq P_{A,s}^n,\ 0 \leq P_{B,s}^n,\ 0 \leq P_{R,m,s}^n,\ \forall m.
\end{equation}

According to~\cite{Ref4}, selecting single relay to forward information is the optimal strategy towards best EE in AF TWMR networks, which depends on two-way data rate requirements and channel conditions. It is assumed that relay $m$ has been selected, i.e., $\rho_{m,s}^n=1$ at relay $m$ and others are equal to $0$. By substituting amplification factor \eqref{Equ2} and achievable data rates \eqref{Equ9}, \eqref{Equ10} into \eqref{Equ24}, the transformed optimization problem can be formulated, as given by
\begin{equation}\label{Equ29}
\begin{aligned}
    &{\rm Min}\ E_{RS} = \\
    &\sigma_n^2\left[\frac{mr_s^n (1 + \alpha_{R,m,s}^n |g_m^n|^2) (1 + \alpha_{R,m,s}^n |h_m^n|^2)}{|h_m^n|^2|g_m^n|^2\alpha_{R,m,s}^n} + \alpha_{R,m,s}^n\right],
\end{aligned}
\end{equation}
\begin{equation}\nonumber
    {\rm S.t.}\ \alpha_{R,m,s}^n > 0,
\end{equation}
where $mr_s^n=\left(2^{\frac{2 \cdot R_{s,n}^1 \cdot N}{W}}-1\right)+\left(2^{\frac{2 \cdot R_{s,n}^2 \cdot N}{W}}-1\right)$.
According to the optimality conditions, $\frac{\partial E_{RS}}{\partial {\alpha_{R,m,s}^n}} = 0$ should hold in any feasible
optimal solution of the problem, based on which the optimal amplification factor can be resolved, as given by
\begin{equation}\label{Equ31}
    \alpha_{R,m,s}^{n^*} = \sqrt{\frac{mr_s^n}{|h_m^n|^2|g_m^n|^2(mr_s^n+1)}}.
\end{equation}
It can be calculated that the second derivative of the objective function $E_{RS}$ in \eqref{Equ29} is larger than zero when $\alpha_{R,m,s}^n > 0$, as given by
\begin{equation}\label{second_derivative}
    \frac{\partial^2 E_{RS}}{\partial {\alpha_{R,m,s}^n}^2 } = \frac{\sigma_n^2 \cdot mr_s^n}{|h_m^n|^2|g_m^n|^2 (\alpha_{R,m,s}^n)^3} > 0.
\end{equation}
Therefore, the optimization problem is convex and $\alpha_{R,m,s}^{n^*}$ is global optimal. As a result, by substituting $\alpha_{R,m,s}^{n^*}$ into $E_{RS}$ the optimal total power consumption can be obtained, as given by
\begin{equation}\label{Equ32}
    P_{min}^{n,m} = \sigma_n^2\left[\frac{mr_s^n}{|h_m^n|^2} + \frac{mr_s^n}{|g_m^n|^2} + 2\sqrt{\frac{mr_s^n(mr_s^n+1)}{|h_m^n|^2|g_m^n|^2}}\right].
\end{equation}

It is known that the optimal relay selection needs to find the relay $d_n^*$ with the minimal optimal total power consumption~\eqref{Equ32} on subcarrier $n$, as written by
\begin{equation}\label{Equ33}
    d_n^* = {\rm arg}\ \min_m\ \left(P_{min}^{n,m}\right).
\end{equation}
It is seen that the optimal relay selection depends on the two directional data rate requirements, which cannot be resolved without the knowledge of services. In order to get the optimal RS first (without knowledge of rate allocation) and reduce complexity, we consider high data rate demands to obtain some approximations. To support high transmission data rate, SE needs to be large enough by some advanced techniques, e.g., higher-order modulation, so that $mr_s^n+1 \approx mr_s^n$ will be satisfied. Therefore, the optimal total power consumption can be approximated, as given by
\begin{equation}\label{Equ34}
    P_{min}^{n,m} \approx mr_s^n\left[\sigma_n^2\left(\frac{1}{|h_m^n|^2} + \frac{1}{|g_m^n|^2} + 2\sqrt{\frac{1}{|h_m^n|^2|g_m^n|^2}}\right)\right].
\end{equation}
As $mr_s^n$ is identical at every relay, the optimal relay selection only depends on two directional channel conditions, which can be expressed as
\begin{equation}\label{Equ35}
    d_n^* = {\rm arg}\ \min_m\ \left[\sigma_n^2\left(\frac{1}{|h_m^n|^2} + \frac{1}{|g_m^n|^2} + 2\sqrt{\frac{1}{|h_m^n|^2|g_m^n|^2}}\right)\right],
\end{equation}
which is called the integrated-noise-channel selection criterion, depending on the comprehensive two-way relay channel conditions. By applying the criterion to all subcarriers, the optimal channel configuration can be obtained, i.e., the connection between subcarrier and relay. The channel coefficients at the selected optimal relay on subcarrier $n$ are denoted by $h_n$ and $g_n$, i.e., $h_n=h_{d_n^*}^n$ and $g_n=g_{d_n^*}^n$. It is assumed that the optimal channel configuration has already been obtained in the rest of the paper.
\subsection{Optimal Subcarrier Assignment}\label{subsec4:2}
Pairing of service and subcarrier is a arduous NP-hard combinatorial optimization problem with enormous binary integer variables, so it is very complicated to find the optimal solution and cannot get its analytical closed-form expression. Genetic Algorithm~(GA), known as a powerful optimization technique based on the principles of genetics and natural selection, can efficiently solve non-differentiable, discontinuous, multi-dimensional constrained, and highly nonlinear problem even with a large number of local optimal points~\cite{Ref30,Ref31}. It is a low-complexity heuristic optimization algorithm that would find the optimal structure by the way of iteration rather than check every possible solution. So far, having a mature theory and perfect legislation, GA has already been extensively applied to NP-hard resource allocation problem in wireless networks~\cite{Ref28,Ref29}.

In this paper, keeping the best solution after selection, the simple ESGA with time-invariant operation is used to solve the subcarrier assignment problem and the efficient design of GA is beyond the scope of our work. As for the implementation, the reciprocal of objective function in \eqref{Equ18} is designed to be the fitness function in ESGA, where the scheme with better energy efficiency will be more likely to be reserved according to roulette selection, followed by time-invariant single-point crossover and mutation.
\subsection{Optimal Rate Allocation}\label{subsec4:3}
It is assumed that one kind of mapping relationship, which may not be optimal, has been obtained during the temporary iteration of ESGA. It is noted that service $s$ is analyzed here and the others can be obtained the same way due to the service independence. Without loss of generality, it is assumed that $N^s(<N)$ subcarriers are assigned to service $s$, forming a subcarrier group, donated by $\mathbb{N}_s$. When subcarrier $n$ is assigned to service $s$, i.e., $n \in \mathbb{N}_s$, it is obtained that under optimal channel configuration $\rho_{{d_n^*},s}^n=1$ and $\rho_{m,s}^n=0,\ \forall m \neq d_n^*$, based on which~\eqref{Equ9} and~\eqref{Equ10} can be simplified, as given by
\begin{equation}\label{Equ37a}
    r_{1,s}^n = \frac{1}{2} \cdot \frac{W}{N} \cdot \log\left(1 + \frac{|h_n|^2|g_n|^2\alpha_{R,{d_n^*},s}^nP_{A,s}^n}{\sigma_n^2 + |g_n|^2\alpha_{R,{d_n^*},s}^n\sigma_n^2}\right),
\end{equation}
\begin{equation}\label{Equ37b}
    r_{2,s}^n = \frac{1}{2} \cdot \frac{W}{N} \cdot \log\left(1 + \frac{|h_n|^2|g_n|^2\alpha_{R,{d_n^*},s}^nP_{B,s}^n}{\sigma_n^2 + |h_n|^2\alpha_{R,{d_n^*},s}^n\sigma_n^2}\right).
\end{equation}
Therefore, the transmission powers on subcarrier $n$ at node A and B thus can be resolved, as given by
\begin{equation}\label{Equ39a}
    P_{A,s}^n = \frac{\left(2^{\frac{2 \cdot r_{1,s}^n \cdot N}{W}} - 1\right)\left(1 + |g_n|^2\alpha_{R,{d_n^*},s}^n\right)\sigma_n^2}{|h_n|^2|g_n|^2\alpha_{R,{d_n^*},s}^n},
\end{equation}
\begin{equation}\label{Equ39b}
 P_{B,s}^n = \frac{\left(2^{\frac{2 \cdot r_{2,s}^n \cdot N}{W}} - 1\right)\left(1 + |h_n|^2\alpha_{R,{d_n^*},s}^n\right)\sigma_n^2}{|h_n|^2|g_n|^2\alpha_{R,{d_n^*},s}^n}.
\end{equation}
In addition, from~\eqref{Equ2} the optimal transmission power at the optimal relay $d_n^*$ on subcarrier $n$ can be expressed as
\begin{equation}\label{Equ41}
    P_{R,{d_n^*},s}^n = \alpha_{R,{d_n^*},s}^n\left(|h_n|^2P_{A,s}^n + |g_n|^2P_{B,s}^n +\sigma_n^2\right).
\end{equation}

And under optimal channel configuration, the optimal amplification factor $\alpha_{R,{d_n^*},s}^n$ in~\eqref{Equ41} is equal to $\alpha_{R,m,s}^{n^*}$ in~\eqref{Equ31} by letting $m = d_n^*$, based on which the objective function of rate allocation problem can be transformed, as given by
\begin{equation}\label{Equ42}
    {\rm Min}\ \sum_{n \in \mathbb{N}_s} \sigma_n^2\left[\frac{mr_s^n}{|h_n|^2} + \frac{mr_s^n}{|g_n|^2} + 2\sqrt{\frac{mr_s^n(mr_s^n+1)}{|h_n|^2|g_n|^2}}\right],
\end{equation}
where $mr_s^n=(2^{\frac{2 \cdot r_{s,n}^1 \cdot N}{W}}-1)+(2^{\frac{2 \cdot r_{s,n}^2 \cdot N}{W}}-1)$. Under high data rate demands situation, $(mr_s^n+1) \approx mr_s^n$ can be obtained similarly, based on which the rate allocation problem is formulated as
\begin{equation}\label{Equ46}
    {\rm Min}\ \sum_{n \in \mathbb{N}_s} \left(mr_s^n \times \eta_n\right),   \tag{\textbf{P3}}
\end{equation}
\begin{equation}\nonumber
    {\rm S.t.}\ \sum_{n \in \mathbb{N}_s} r_{1,s}^n = R_s^1,\ \sum_{n \in \mathbb{N}_s} r_{2,s}^n = R_s^2,
\end{equation}
\begin{equation}\nonumber
    P_{A,s}^n \geq 0,\ P_{B,s}^n \geq 0,\ P_{R,{d_n^*},s}^n \geq 0,
\end{equation}
where $\eta_n=\left[\sigma_n^2(\frac{1}{|h_n|^2} + \frac{1}{|g_n|^2} + 2\sqrt{\frac{1}{|h_n|^2|g_n|^2}})\right]$.
It is noted that $r_{1,s}^n=0$ and $r_{2,s}^n=0$ if subcarrier $n$ is not assigned to service $s$.

The Lagrange function that combines the objective function and data rate constraints of this optimization problem is obtained, as given by
\begin{equation}\label{Equ50}
\begin{aligned}
    \mathcal{L} = & \sum_{n \in \mathbb{N}_s} \left(mr_s^n \times \eta_n\right) - \lambda_1 \left( \sum_{n \in \mathbb{N}_s} r_{1,s}^n - R_s^1 \right) \\
    &- \lambda_2 \left( \sum_{n \in \mathbb{N}_s} r_{2,s}^n - R_s^2 \right),
\end{aligned}
\end{equation}
where $\lambda_1$ and $\lambda_2$ are Lagrange multiplier coefficients. According to Karush-Kuhn-Tucker~(KKT) conditions, $\frac{\partial \mathcal{L}}{\partial {r_{1,s}^n}} = 0$ and $\frac{\partial \mathcal{L}}{\partial {r_{2,s}^n}} = 0$ should hold simultaneously for all allocated subcarrier $n$. Therefore, the following relationship must be verified, as given by
\begin{equation}\label{Equ51a}
    r_{1,s}^n = \frac{1}{2} \cdot \frac{W}{N} \cdot \log\left(\frac{\lambda_1}{\eta_n \cdot 2 \cdot ln2}\right),\ n \in \mathbb{N}_s,
\end{equation}
\begin{equation}\label{Equ51b}
    r_{2,s}^n = \frac{1}{2} \cdot \frac{W}{N} \cdot \log\left(\frac{\lambda_2}{\eta_n \cdot 2 \cdot ln2}\right),\ n \in \mathbb{N}_s.
\end{equation}
By substituting \eqref{Equ51a} and \eqref{Equ51b} into the sum-rate constraints in \eqref{Equ46}, the following relationship must be satisfied
\begin{equation}\label{Equ52}
\begin{aligned}
    &\sum_{n \in \mathbb{N}_s} \log\left(\frac{B_1}{\eta_n}\right)=\frac{2 \cdot R_s^1 \cdot N}{W},\\
    &\sum_{n \in \mathbb{N}_s} \log\left(\frac{B_2}{\eta_n}\right)=\frac{2 \cdot R_s^2 \cdot N}{W}.
\end{aligned}
\end{equation}
where $B_1=\frac{\lambda_1}{2 \cdot ln2}$ and $B_2=\frac{\lambda_2}{2 \cdot ln2}$. Thus the water level can be resolved, as written by
\begin{equation}\label{Equ53}
    B_1=\sqrt[N_s]{2^{\frac{2 \cdot R_s^1 \cdot N}{W}} \prod_{n \in \mathbb{N}_s} \eta_n},\ B_2=\sqrt[N_s]{2^{\frac{2 \cdot R_s^2 \cdot N}{W}} \prod_{n \in \mathbb{N}_s} \eta_n}.
\end{equation}
The optimal rate allocation for service $s$ can be obtained consequently, as formulated by
\begin{equation}\label{Equ54}
\begin{aligned}
    &r_{1,s}^n = \frac{1}{2} \cdot \frac{W}{N} \cdot \log \left(\frac{B_1}{\eta_n}\right),\ n \in \mathbb{N}_s,\\
    &r_{2,s}^n = \frac{1}{2} \cdot \frac{W}{N} \cdot \log \left(\frac{B_2}{\eta_n}\right),\ n \in \mathbb{N}_s.
\end{aligned}
\end{equation}
If any calculated data rate is negative, which is impossible in practice, it is inferred that the subcarrier assignment is wrong, i.e., some subcarriers, whose noise-channel coefficient is larger than the water level, are selected. These subcarriers then should be excluded and then recalculate the water level again until all the obtained data rates are positive.
\section{Equivalent Optimization Objective Function}\label{sec5}
From what we have discussed above, no matter in what situation, the optimal channel configuration will be conducted first regardless of subcarrier assignment and its corresponding rate allocation. Therefore, with optimal channel configuration, i.e., optimal single relay has been selected on each subcarrier, according to~\eqref{Equ46}, the objective function of~\eqref{Equ18} under high data rate demand can be transformed, as given by
\begin{equation}\label{Equi_Optim}
    E^h_{sum} = \sum_{s=1}^S \sum_{n=1}^N C_s^n \left(mr_s^n \times \eta_n\right).
\end{equation}
If do not distinguish different services and treat them as a whole service loading all data rate demands, the optimal rate allocation can be fulfilled through the process mentioned in Section~\ref{subsec4:3}. Using Lagrange multiplier method or convex optimization theory, the optimal water level in two directions are resolved, as given by
\begin{equation}\nonumber
    {B_1}^*=\sqrt[N]{2^{\frac{2 \cdot \sum_{s=1}^S R_s^1 \cdot N}{W}} \prod_{n=1}^N \eta_n},\ {B_2}^*=\sqrt[N]{2^{\frac{2 \cdot \sum_{s=1}^S R_s^2 \cdot N}{W}} \prod_{n=1}^N \eta_n}.
\end{equation}
Therefore, according to~\eqref{Equ54}, the optimal rate allocation on every subcarrier $n$, i.e., ${r_1^n}^*$ and ${r_2^n}^*$, can be obtained. Since hybrid services cannot share a certain subcarrier, the single virtual service case we introduced is the ideal situation with best fixed EE performance. This certain upper bound cannot be exceeded for hybrid services in practice. It can be found that minimizing the objective function~\eqref{Equi_Optim} means reducing the difference between $E^h_{sum}$ and the fixed ideal EE in single virtual service case.

From the perspective of the subcarrier, the rate allocation result for hybrid services is actually a series of data rates on each subcarrier, denoted by $r_1^n$ and $r_2^n$, where
\begin{equation}\nonumber
    r_1^n = r_{1,s}^n,\  r_2^n = r_{2,s}^n,\ \left(s = {\rm arg}\ \{C_s^n=1\}\right).
\end{equation}
It is seen that $r_1^n$ and $r_2^n$ have contained the knowledge of $C_s^n$, therefore, given any rate allocation result, the EE difference of the two cases, referred as the new equivalent optimization objective function, can be formulated, as given by
\begin{equation}\label{D_value}
    \triangle E^h_{sum} = \sum_{n=1}^N \left[\left({mr^n - mr^n}^*\right) \times \eta_n\right],
\end{equation}
where $mr^n=\left(2^{\frac{2 \cdot r_{1}^n \cdot N}{W}}-1\right)+\left(2^{\frac{2 \cdot r_{2}^n \cdot N}{W}}-1\right),\ {mr^n}^*=\left(2^{\frac{2 \cdot {r_{1}^n}^* \cdot N}{W}}-1\right)+\left(2^{\frac{2 \cdot {r_{2}^n}^* \cdot N}{W}}-1\right)$.
Due to the independence between downlink and uplink, problem~(\ref{D_value}) can be separated into two similar subfunctions, where the downlink case, for example, is presented with mathematical manipulation, as given by
\begin{equation}\nonumber
\begin{aligned}
    \triangle E^{h,d}_{sum} & = \sum_{n=1}^N \left[\left(2^{\frac{2 \cdot r_{1}^n \cdot N}{W}}-2^{\frac{2 \cdot {r_{1}^n}^* \cdot N}{W}}\right) \times \eta_n\right] \\
    & = \sum_{n=1}^N \left[\left(2^{\frac{2 \cdot (r_{1}^n-{r_{1}^n}^*) \cdot N}{W}}-1\right) \cdot 2^{\frac{2 \cdot {r_{1}^n}^* \cdot N}{W}} \cdot \eta_n\right].
\end{aligned}
\end{equation}
It is obvious that~\eqref{D_value} can reach minimum if and only if the two subfunctions, i.e., downlink and uplink case, become minimum at the same time. The downlink case is studied here and the uplink case can be obtained similarly. It is attained that ${B_1}^*=2^{\frac{2 \cdot {r_{1}^n}^* \cdot N}{W}} \cdot \eta_n$ and ${B_2}^*=2^{\frac{2 \cdot {r_{2}^n}^* \cdot N}{W}} \cdot \eta_n$ because ${r_1^n}^* = \frac{1}{2} \cdot \frac{W}{N}\log \left(\frac{{B_1}^*}{\eta_n}\right)$ and ${r_2^n}^* = \frac{1}{2} \cdot \frac{W}{N} \log \left(\frac{{B_2}^*}{\eta_n}\right)$, based on which removing the constant terms, the equivalent optimization objective function including uplink and downlink case can be obtained, as given by
\begin{equation}\label{EOC}
    \triangle E^{h,a}_{sum} = {B_1}^* \cdot \sum_{n=1}^N 2^{\frac{2 \cdot \triangle r_1^n \cdot N}{W}} + {B_2}^* \cdot \sum_{n=1}^N 2^{\frac{2 \cdot \triangle r_2^n \cdot N}{W}},
\end{equation}
where $\triangle r_1^n = r_{1}^n-{r_{1}^n}^*,\ \triangle r_2^n = r_{2}^n-{r_{2}^n}^*$.
\eqref{EOC} is called the equivalent optimization objective function, which can be set as the simple evaluating index in ESGA. It is known that the more minimal the objective function~\eqref{EOC} is, the better the rate allocation result will be.
\section{Optimization Algorithm}\label{sec6}
In the algorithm, the optimal relay for each subcarrier is first selected according to the integrated-noise-channel selection criterion we proposed. The subcarrier assignment sequence is then initialized randomly with the constraints of $\{C_s^n\}$ for all services. The optimal rate allocation can be conducted according to~\eqref{Equ54} and its fitness can be calculated via the equivalent optimization function~\eqref{EOC}. Through subcarrier assignment iteration of ESGA, the optimal sequence will be obtained by selection, crossover and mutation until end condition is satisfied. Finally, optimal EE solution is attained with best subcarrier assignment and rate allocation. The description of the algorithm is presented in~\textbf{Algorithm \ref{Algorithm_MTWF1}}.
\begin{algorithm}[!t]
\footnotesize
\caption{The mapped two-way water filling allocation}
\label{Algorithm_MTWF1}
\begin{algorithmic}[1]
    \REQUIRE Chromosome number in ESGA: $Popsize$, iteration number in ESGA: $g_c$, service number: $S$, channel conditions: $\{h_m^n,\ g_m^n\}$, QoS requirements, and bursty traffic features;
    \STATE Obtain sum-rate constraints from the QoS requirements through equivalent queue analysis for all services;
    \STATE According to the integrated-noise-channel selection criterion, the optimal relay for each subcarrier is selected and the optimal channel conditions are recorded;
    \STATE Calculate the comprehensive two-way channel coefficients for each subcarrier;
    \FOR{$i=1:Popsize$}
        \STATE Initialize the subcarrier assignment sequence satisfying the $\{C_s^n\}$ constraints of each service for chromosome $i$;
    \ENDFOR
    \FOR{$g=1:g_c$}
        \FOR{$i=1:Popsize$}
            \FOR{$j=1:S$}
                \STATE Resolve water level according to~\eqref{Equ53} for service $j$; \label{Cal_water}
                \STATE Conduct the rate allocation and obtain the data rate ${r_1}$ and ${r_2}$ according to water level for service $j$;
                \IF{$\forall {r_1} < 0$,\ $\forall {r_2} < 0$}
                    \STATE let these negative data rates be 0 and exclude these subcarriers. Turn to Step~\ref{Cal_water};
                \ENDIF
            \ENDFOR
            \STATE Calculate the fitness according to the equivalent optimization function~(\ref{EOC}) as evaluating indicator;
        \ENDFOR
        \STATE Select and record the best assignment scheme;
        \STATE Obtain the next generation by crossover and mutation;
    \ENDFOR
\end{algorithmic}
\end{algorithm}
\subsection{Optimality Analysis}\label{subsec6:1}
It is known that the convergence property is the basic theory of GA, which has been widely studied and many mature results can be found in the existing literatures~\cite{Ref30,Ref31,Ref32}. The building block hypothesis shows that the solutions with large fitness can combine with each other and then create higher fitness solution, which can eventually generate the global optimal solution~\cite{Ref31}; in other words, GA has the ability to find the global optimal solution. And the schema theorem ensures that the number of the better solution in GA is exponential growth, which satisfies the necessary condition of finding the global optimal solution~\cite{Ref30}. In~\cite{Ref32}, it is proved by homogeneous finite Markov chain analysis that ESGA, which always keeps the best solution in the population after selection, will converge to the global optimum as the iteration number increases.


It is known that our scheme contains three parts, including relay selection, subcarrier assignment, and rate allocation. As we solved optimal relay selection and rate allocation problems by convex optimization methods, the optimality of the solution about these two aspects can be ensured, respectively. It is obtained that under high data rate demand, relay selection has nothing to do with subcarrier assignment and rate allocation, i.e., the optimality of relay selection is independent with the latter two parts. Given any result of subcarrier assignment, the global optimal rate allocation is fixed, which can be obtained through the proposed two-way water filling principle; in other words, each sample of the solution in ESGA iteration is theoretical optimal in terms of rate allocation. Therefore, the optimality of the searched solution only depends on subcarrier assignment and its solving method. Since ESGA used in this paper can converge to the global optimum, we can ensure the optimal solution we obtained about subcarrier assignment and rate allocation is the global optimum with enough iteration. Combined with the independent global optimality of relay selection, the global optimal solution, resolved by our proposed scheme, can be guaranteed.
\subsection{Complexity Analysis}\label{subsec6:2}
In this subsection, we will analyse the computational complexity of our proposed scheme in units of computer calculation, including multiplication, division, addition, subtraction, and judgement etc. $Popsize$ and $g_c$ are the chromosome number and iteration number in ESGA, which indicate the scale of the algorithm together with $S$, $N$, and $M$. Instead of expressing the complexity of the whole algorithm, we first get the complexity of each step and then combine them to the final one.

The sum-rate requirement needs to be calculated for every service, therefore the complexity is $\mathcal{O}(S)$. During optimal relay selection, $10 \hspace{-0.5mm} \times \hspace{-0.5mm} M \hspace{-0.5mm} \times \hspace{-0.5mm} N$ mathematical operations and $(M \hspace{-0.5mm} - \hspace{-0.5mm} 1) \hspace{-0.5mm} \times \hspace{-0.5mm} N$ judgements are required, thereby its complexity is $\mathcal{O}(M \hspace{-1mm} \times \hspace{-1mm} N)$. In terms of the two-way water filling, for each service the complexity $\mathcal{O}(N)$ is needed for water level and optimal rate calculation. As for the computational complexity of ESGA, $\mathcal{O}(N \hspace{-1mm} \times \hspace{-1mm} Popsize)$ is for initialization; $\mathcal{O}(N \hspace{-1mm} \times \hspace{-1mm} S)$ is for fitness computation; $\mathcal{O}(Popsize^2)$ is for selection; $\mathcal{O}(N \hspace{-1mm} \times \hspace{-1mm} S \hspace{-1mm} \times \hspace{-1mm} Popsize)$ is for crossover; $\mathcal{O}(N \hspace{-1mm} \times \hspace{-1mm} Popsize)$ is for mutation. According to the procedure in~\textbf{Algorithm \ref{Algorithm_MTWF1}}, the computational complexity of our scheme, donated by $\mathcal{O}_w$, can be calculated, as given by
\begin{equation}\label{complexity}
\begin{aligned}
    \mathcal{O}_w &= \{[S \hspace{-1mm} \times \hspace{-1mm} \mathcal{O}(N) \hspace{-1mm} + \hspace{-1mm} \mathcal{O}(N \hspace{-1mm} \times \hspace{-1mm} S)] \hspace{-1mm} \times \hspace{-1mm} Popsize \hspace{-1mm} + \hspace{-1mm} \mathcal{O}(Popsize^2) \hspace{-1mm} \\
    & \ \ \ + \hspace{-1mm} \mathcal{O}(N \hspace{-1mm} \times \hspace{-1mm} S \hspace{-1mm} \times \hspace{-1mm} Popsize)\} \hspace{-1mm} \times \hspace{-1mm} g_c \hspace{-1mm} + \hspace{-1mm} \mathcal{O}(S) \hspace{-1mm} + \hspace{-1mm} \mathcal{O}(M \hspace{-1mm} \times \hspace{-1mm} N) \\
    & = \mathcal{O}(N \hspace{-1mm} \times \hspace{-1mm} S \hspace{-1mm} \times \hspace{-1mm} Popsize \hspace{-1mm} \times \hspace{-1mm} g_c \hspace{-1mm} + \hspace{-1mm} Popsize^2 \hspace{-1mm} \times \hspace{-1mm} g_c \hspace{-1mm} + \hspace{-1mm} M \hspace{-1mm} \times \hspace{-1mm} N),
\end{aligned}
\end{equation}
which is the polynomial computational complexity w.r.t $Popsize$, $N$, $S$, $M$, and $g_c$.
\section{Simulation results}\label{sec7}
In order to evaluate the potential profit of our proposed scheme in TWMR OFDM networks, Mapped Average~(MA), Mapped Water Filling~(MWF), and Mapped GA~(MGA) allocation algorithm are simulated for comparison, where MA distributes the data rate demands to each subcarrier equally; MWF fulfills the rate allocation by traditional water filling scheme; MGA applies ESGA to achieve both subcarrier assignment and rate allocation. In simulations, the square of channel coefficient is generated directly, and the power spectral density of noise at all nodes is assumed to be equal to 1. To further control channel difference, by introducing channel difference coefficient $PLC$, $|h_m^n|^2$ and $|g_m^n|^2$ can be expressed as
\begin{equation}\nonumber
    |h_m^n|^2=h_r^{PLC},\ |g_m^n|^2=g_r^{PLC},\ \forall n,m.
\end{equation}
where $h_r$ and $g_r$ are both random numbers.
\subsection{Symmetric And Asymmetric Traffic}\label{subsec7:1}
\begin{figure}[!t]
\centering
\includegraphics[scale=0.3]{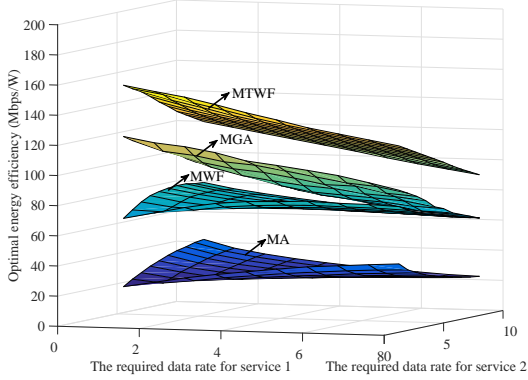}
\vspace{-3mm}
\caption{The optimal EE for different required data rates (bps/Hz) in two directions with symmetric traffic.}
\label{R_3D}
\end{figure}

Fig.~\ref{R_3D} shows the optimal EE on different required data rates for 2 services where symmetric traffic in two directions is assumed, and 16 subcarriers and 6 relays are contained. It is obvious that MTWF has the best EE performance due to the optimal rate allocation compared with MA, MWF, and MGA. It is because that MA does not optimize the rate allocation completely, while MWF only contains respective direction optimization rather than the joint consideration at the same time. However MGA, also optimal theoretically, cannot reach the best EE performance like MTWF due to the randomness of ESGA in rate allocation. It is also observed that the EE of MTWF and MGA will decline with the increase of the required data rates on each service. It is because that in order to support higher data rates, the transmission power needs to increase exponentially according to Shannon capacity formula.

Unlike MTWF and MGA, the EE of MA and MWF will first rise and then decrease with the increase of required data rates. The channel conditions play the more important role in optimizing rate allocation when the required data rate is low, because the allocated transmission power in this region has the same magnitude as the channel difference or even smaller. It is known that MA does not consider the channel conditions, and MWF only optimizes respective directional transmission, so that their EE performances are poor in low data rate region. However, the effect of channel difference becomes weaker with greater allocated power, based on which the higher EE appears with the increase of required data rates. Specially, the power allocation result of MGA and MWF is the same as MTWF when the data rate demand turns into infinite. Therefore, the EE performance is first improved due to the reduction of channel difference and then decreases since the transmission power increases exponentially to support higher rate.

\begin{figure}[!t]
\centering
\includegraphics[scale=0.3]{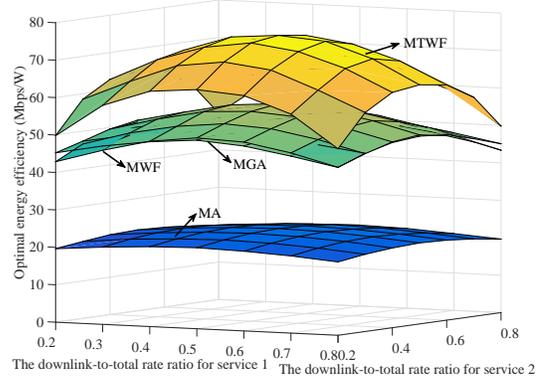}
\vspace{-3mm}
\caption{The optimal EE for different downlink-to-total data rate ratios ${R_1}/({R_1+R_2})$ for each service.}
\label{Ratio_Two_Direction}
\end{figure}

Fig.~\ref{Ratio_Two_Direction} plots the optimal EE under asymmetric traffic in two directions where 16 subcarriers, 2 services and 6 relays are included. It is noted that the imbalance traffic degree first relieves and then becomes severer with the increase of the downlink-to-total rate ratio. Specifically, the balance traffic is achieved when this ratio is equal to 0.5. It is shown that MTWF can obtain the best EE performance, while MA is the worst scheme. The EE of all schemes will decrease with the enhancement of asymmetric traffic. The explanation is that the downlink and uplink are associated in two-way relay networks and the EE performance depends on the worse directional transmission that loads higher data rate demand.

\begin{figure}[!t]
\centering
\includegraphics[scale=0.3]{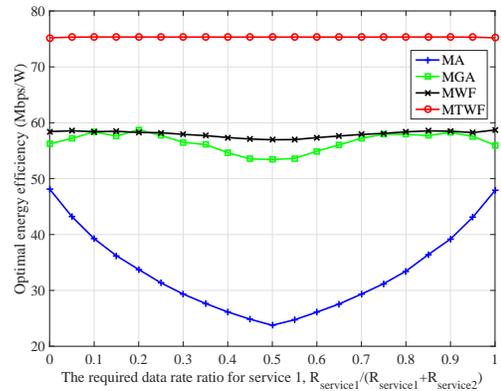}
\vspace{-3mm}
\caption{The optimal EE for different required data rate ratios of services with symmetric traffic in two directions.}
\label{Ratio_Service}
\end{figure}

Fig.~\ref{Ratio_Service} shows the optimal EE under asymmetric traffic for 2 services but with symmetric traffic in two directions, where 16 subcarriers and 6 relays are involved. It is observed that by assigning the optimal rate on each subcarrier to suitable service, MTWF has the identical EE under asymmetric traffic, because its rate allocation result can get almost the same output as single virtual service case, that is the insurmountable upper bound. Similarly, by the rational assignment of subcarriers, e.g., more subcarriers will be assigned to the service loading more traffic, MWF can almost maintain its performance for the asymmetric traffic. However, the EE of MA becomes better with the increase of asymmetric traffic. The reason is that with higher data rate for one service, more subcarriers will be assigned to this service to avoid exponentially increased transmission power, and the subcarriers for the other service will become less consequently. Higher data rate for one service and less subcarriers for other service will decrease the channel difference effect, i.e., will increase the EE as the total data traffic demand is fixed. It is known that the EE of MGA depends on the probability of finding the optimum, i.e., the number of feasible solutions. The larger the number is, the worse the EE will be. It is known that the number of the feasible results will first decrease and then rise, causing the curve tendency of MGA.
\subsection{The Effect Of System Parameter}\label{subsec7:2}
\begin{figure}[!t]
\centering
\includegraphics[scale=0.3]{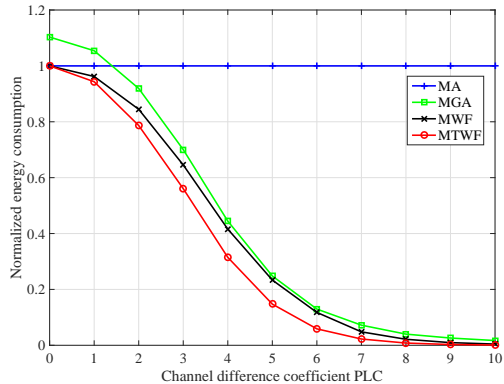}
\vspace{-3mm}
\caption{Normalized energy consumption under different channel difference coefficients $PLC$ with symmetric traffic.}
\label{picChannel}
\end{figure}

\begin{figure}[!t]
\centering
\includegraphics[scale=0.3]{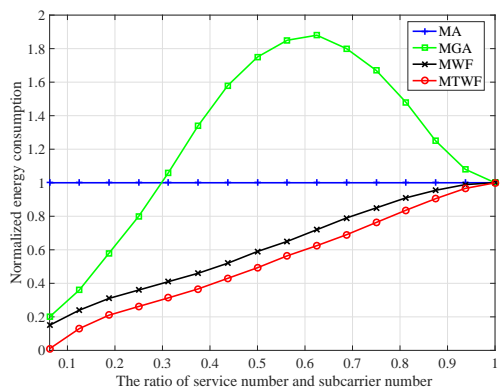}
\vspace{-3mm}
\caption{Normalized energy for different ratios of service number and subcarrier number with symmetric traffic.}
\label{picService}
\end{figure}

In Fig.~\ref{picChannel} and Fig.~\ref{picService}, the vertical axis is the normalized energy consumption that can be interpreted as the optimization degree against to MA. From Fig.~\ref{picChannel}, it is seen that MTWF has the best EE compared with other schemes. The reason is that MA does not consider and utilize the channel conditions, which play a more important role in rate allocation with the increase of $PLC$, which causes EE deterioration. Though MWF will use the channel information, it conducts optimization only concentrating on respective directions rather than joint consideration, thereby it is much better than MA but a little worse than MTWF. It is observed that MTWF, MWF, and MA get the same EE performance when $PLC=0$, i.e., all the channel conditions are the same specially, because each service distributes data rate demand to all subcarriers equally in MTWF and MWF. There is a stable gap on EE between MTWF and MGA due to the MGA's randomness of find the optimal solution.

Fig.~\ref{picService} plots the normalized energy consumption for different service numbers with fixed overall data traffic demand, where 16 subcarriers and 6 relays are involved. It is observed that with the increase of service number, the performance of MWF and MTWF gets worse while MGA's performance first ascends and then declines. When service number and subcarrier number tend to the same in the end, only one subcarrier will be assigned to each service. As for MWF and MTWF, due to the increase of service number, the subcarriers that can be allocated to a single service are reduced, i.e., the parallel channels are reduced, which causes the decrease of diversity degree. The performance reduction of single service will further lead to the EE falling for the whole scheme due to the service independence. It is known that the probability that MGA can find the optimal solution is determined by the feasible solution number. The more the feasible solutions are, the less the probability is, i.e., the larger the normalized energy consumption will be. According to permutation and combination, the mapping patterns first rises and then decreases with the service number increment, which is identical to the curve tendency of MGA.

\begin{figure}[!t]
\centering
\includegraphics[scale=0.3]{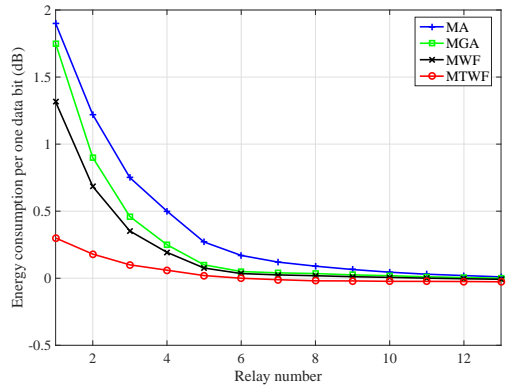}
\vspace{-3mm}
\caption{The energy consumption per one data bit under different relay numbers with symmetric traffic.}
\label{picRelay}
\end{figure}

Fig.~\ref{picRelay} shows the relay's influence in terms of energy consumption per unit data bit, presented in the form of decibel. For all the four schemes, this index will decrease gradually with the increase of relay number, indicating that relay can bring diversity gain to improve the energy performance. However, the gain becomes smaller, i.e., the gain brought by each additional relay will be less. Moreover, the performance gap between the four schemes will decrease sharply, especially the performances are almost the same after relay number exceeding a certain threshold.

\begin{figure}[!t]
\centering
\includegraphics[scale=0.3]{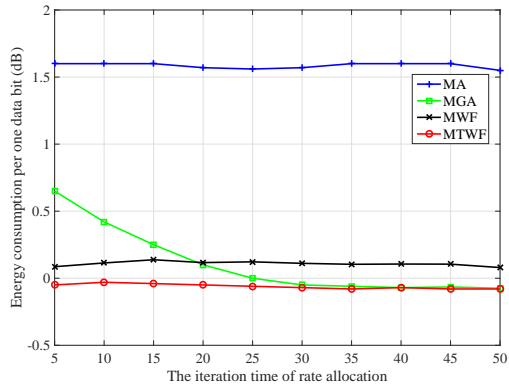}
\vspace{-3mm}
\caption{Energy consumption per one data bit for iteration time of rate allocation with symmetric traffic.}
\label{picInloop}
\end{figure}

\begin{figure}[!t]
\centering
\includegraphics[scale=0.3]{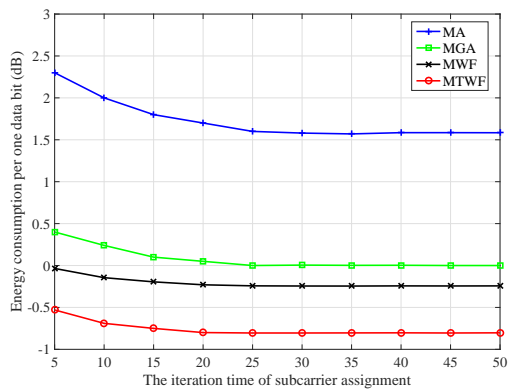}
\vspace{-3mm}
\caption{Energy consumption per one data bit for iteration time of subcarrier assignment with symmetric traffic.}
\label{picOutloop}
\end{figure}

Fig.~\ref{picInloop} shows that the energy consumption per unit bit of MGA is reduced and reaches the same as MTWF with the increase of rate allocation iteration. It is because that MGA will find better solution, while MA, MWF, and MTWF almost do not change due to the irrelevance with this iteration. Specially when the iteration exceeds a threshold, MGA can find the optimal solution like MTWF and the performance becomes stable. Fig.~\ref{picOutloop} indicates that the energy consumption per unit bit for the four schemes will drop as the subcarrier assignment iteration increases. When the iteration exceeds a threshold, the performance becomes stable because the optimal solution has already been found. It is observed that the algorithm tends to converge with evolution and iteration the same as our theoretical analysis.
\section{Conclusion}\label{sec8}
In this paper, we studied the energy-efficient resource allocation for hybrid bursty services with QoS requirements in TWMR OFDM networks. The strictly proved equivalent homogeneous Poisson process was developed to analyze the bursty traffic in terms of average performance, base on which the QoS requirements were converted into the sum-rate constraints. Without avoiding the non-convex NP-hard combinatorial optimization problem, after obtaining the optimal relay configuration by some approximations, subcarrier assignment is fulfilled by ESGA and we proposed two-way water filling principle to conduct optimal rate allocation for each service. To reduce the complexity of ESGA, the equivalent objective function was proposed which can be set as the simple evaluating index. Simulation results showed the superiority and convergence of our scheme and we analyzed the performance on different scenarios.

\appendices
\section{Proof of the average performance equivalent}\label{Apx1}
With the same data packet departure (that is homogeneous Poisson process), there are two kinds of data packet arrival, which are equivalent about the average performance, i.e., delay in this paper, as given by

\vspace{1mm}

{Homogeneous Poisson arrival:} The data packets arrive with a fixed rate $\lambda$ with a period of time $T$.

\vspace{0.5mm}

{Heterogeneous Poisson arrival:} The data packets arrive with a changed rate $\lambda(t)$ w.r.t time $t$ within a period of time $T$, and the following constraint is satisfied, as given by
\begin{equation}\label{Equi_Con}
    \lambda \cdot T = \int_0^T \lambda(t) \cdot dt.
\end{equation}

\vspace{1mm}

Since the departure process is identical, the equivalence of the two queue systems can be ensured if the equivalence of the two arrival processes can be proved within time duration $T$. Observing the whole queue system at time $T$, it is seen that the waiting time, depending on the state of queue system, will be same for different queue systems as long as the probability of queue length is identical. It is noted that only same average performance at time $T$ is guaranteed, but not the transient performance or any time before $T$.

As for homogeneous poisson arrival, the probability of arriving $k$ data packets within time $T$ can be expressed as
\begin{equation}\nonumber
    P_k^1(T)=\frac{(\lambda \cdot T)^k}{k!}e^{-\lambda \cdot T}.
\end{equation}

Similarly, for heterogeneous poisson arrival, this probability of arriving $k$ data packets can be obtained, as given by
\begin{equation}\nonumber
    P_k^2(T)=\frac{\left(\int_0^T\lambda(t) \cdot dt\right)^k}{k!}e^{-\int_0^T\lambda(t) \cdot dt}.
\end{equation}
According to \eqref{Equi_Con}, $P_k^1(T)=P_k^2(T)$ is obtained which indicates that the probability of arriving $k$ data packets within time $T$ is identical. It is obvious that the probability of leaving $q$ data packets within time $T$ is identical too due to the same departure process, resulting in the same state of queue system. Therefore, it is proved that the two kinds of process are equivalent if the constraint~\eqref{Equi_Con} is satisfied.

The average time duration can be written as $T_1^s+\frac{1}{\Lambda_1^s}$ for service $s$ in downlink from a burst arrival to another burst arrival periodically. The burst data packet arrival is a heterogeneous poisson process, thus the following relationship can be obtained with the arrival rate $\lambda_1^s$ during bursty duration while $0$ during bursty interval, as given by
\begin{equation}\nonumber
    \int_0^{T_1^s+\frac{1}{\Lambda_1^s}} \lambda(t) \cdot dt = \int_0^{T_1^s} \lambda_1^s \cdot dt + \int_{T_1^s}^{T_1^s+\frac{1}{\Lambda_1^s}} 0 \cdot dt=\lambda_1^s \cdot T_1^s.
\end{equation}

Since ${\lambda_1^s}^*=\frac{\lambda_1^s \cdot T_1^s}{{T_1^s+\frac{1}{\Lambda_1^s}}}$, the new homogeneous poisson arrival with the parameter of ${\lambda_1^s}^*$ can meet the constraint~\eqref{Equi_Con} due to the following relationship, as given by
\begin{equation}\nonumber
    {\lambda_1^s}^*\left({T_1^s+\frac{1}{\Lambda_1^s}}\right)= \lambda_1^s \cdot T_1^s =\int_0^{T_1^s+\frac{1}{\Lambda_1^s}} \lambda(t) \cdot dt.
\end{equation}

Therefore, it is completely proved that it is equivalent on average performance for the bursty arrival and the homogeneous poisson arrival with the parameter of ${\lambda_1^s}^*$.
\end{document}